\documentstyle[12pt]{article}

\begin{document}

\begin{center}
{\large {\bf 
A multistream model for quantum plasmas}}
\vskip 5mm
 
{ F.~Haas$^1$, G. Manfredi$^2$} \\ Laboratoire de 
Physique des Milieux Ionis\'es, Universit\'e Henri Poincar\'e, \\
BP 239, 54506 Vandoeuvre-les-Nancy, France 

{ M. Feix} \\
Subatech, Ecole des Mines de Nantes, BP 20722, 44397 Nantes Cedex 3, France
\end{center}
\vskip 5mm

{\bf Abstract}

The dynamics of a quantum plasma can be described 
self-consistently by the nonlinear 
Schr\"odinger-Poisson system. 
Here, we consider a multistream model
representing a statistical mixture of $N$ pure states, each described by a 
wavefunction. The one-stream and two-stream cases are investigated.
We derive the dispersion relation for the two-stream instability 
and show that a new, purely quantum, branch appears.
Numerical simulations of the complete Schr\"odinger-Poisson system
confirm the linear analysis, and provide further results 
in the strongly nonlinear regime. 
The stationary states of the Schr\"odinger-Poisson 
system are also investigated. These can be viewed as the 
quantum mechanical counterpart of the classical Bernstein-Greene-Kruskal modes, and
are described by a set of coupled nonlinear differential equations for the 
electrostatic potential and the stream amplitudes.
\vskip 5mm

$^1$ Fernando.Haas@lpmi.uhp-nancy.fr \\
$^2$ Giovanni.Manfredi@lpmi.uhp-nancy.fr

\newpage

\section{Introduction}

The great degree of miniaturization of today's electronic 
components is such that the de Broglie wavelength of the charge carriers is 
frequently comparable to the dimensions of the system. Hence, 
quantum mechanical effects (e.g. tunneling)
are expected to play a central role in the behavior of 
electronic components to be constructed in the next years. 
In order to describe these quantum effects, it is unlikely that 
classical transport models will be sufficient. Quantum  
transport equations, such as the Schr\"odinger-Poisson or the 
Wigner-Poisson systems \cite{Arnold}--\cite{Luscombe}, 
will therefore be a necessary tool 
in order to understand the basic properties of these physical systems. 

In the present paper, we consider a one-dimensional quantum plasma,
where the electrons are described by a 
statistical mixture of $N$ pure states,
with each wavefunction $\psi_i$ obeying the
Schr\"odinger-Poisson system
\begin{eqnarray}
\label{schro}
i\hbar\frac{\partial\psi_i}{\partial\,t} &=&
- \frac{\hbar^2}{2M}\frac{\partial^{2}\psi_i}{\partial x^2} 
- e\phi\psi_i~, ~~~ i = 1,...,N \\
\label{poisson}
\frac{\partial^2 \phi}{\partial x^2} &=& 
\frac{e}{\varepsilon_0}(\sum_{i=1}^{N}|\psi_{i}|^2
- n_{0}) \,,
\end{eqnarray}
where $\phi(x,t)$ is the electrostatic potential.
Electrons have mass $M$ and charge $- e$, and are globally neutralized by
a fixed ion background with density $n_0$. Finally, we 
assume periodic boundary conditions, with spatial period equal to $L$. 

The system of Eqs. (\ref{schro})--(\ref{poisson}) takes into account diffraction, 
which is the most prominent quantum effect, but neglects 
dissipation, spin and relativistic corrections. These effects may 
be important in more realistic models for small semiconductor 
devices. Nevertheless, it is useful to consider 
simplified models that capture the main features of quantum 
plasmas. Indeed, Eqs. (\ref{schro})--(\ref{poisson}) are 
sufficiently rich to display a wide variety of behaviors, as it will 
be seen in the rest of this work. At the same time, this model is still 
amenable to analytic and numerical treatment. 

A physically equivalent 
approach would consist in considering a 
Wigner function describing the same mixture. The Wigner function approach
is a reformulation of quantum mechanics in the classical phase space language
\cite{Wigner}--\cite{Tatarski}.
The price to pay for this otherwise appealing formalism, is that Wigner functions
can take negative values, and cannot therefore be regarded as true probability 
distributions. However,  
both for the analytical and the numerical treatment of the problems 
of interest in this paper, the 
Schr\"odinger-Poisson formalism is more appropriate. This is 
particularly true for the numerical simulations, since the Wigner 
formalism is cast into a two-dimensional phase space, whilst 
the Schr\"odinger-Poisson model only requires the discretization of the 
one-dimensional configuration space.
Of course, if the number $N$ of streams is large, the numerical cost for 
the description of the system of Eqs. 
(\ref{schro})-(\ref{poisson}) is also considerable. 
Nevertheless, interesting physical phenomena (such as instabilities)
can take place even with a few streams ($N=1$ or 2), 
as it will appear in the rest of this work.  
More subtle effects (Landau damping, for instance) would probably require a 
larger, although hopefully not prohibitive, number of streams 
($N \simeq 20-30$).

For the analytical study,
the hydrodynamic formulation of the Schr\"odinger-Poisson 
system is particularly convenient, since it makes direct use of macroscopic plasma
quantities, such as density and average velocity. Moreover, it
enables one to perform straightforward perturbation calculations in the
same fashion as in the classical case. Let us introduce the amplitude
$A_i$ and the phase $S_i$ associated to the pure state $\psi_i$ according to
\begin{equation}
\label{as}
\psi_i = A_{i}\exp({i\,S_{i}/\hbar}) \,.
\end{equation}
Both $A_i$ and $S_i$ are defined as real quantities. The density $n_i$ and the
velocity $u_i$ of the $i-$th stream of the plasma are given by
\begin{equation}
\label{nu}
n_i = A_{i}^2 \,,\qquad  u_i = 
\frac{1}{M}\frac{\partial\,S_i}{\partial\,x} \,.
\end{equation}
Introducing Eqs. (\ref{as})--(\ref{nu}) into Eqs. (\ref{schro})--(\ref{poisson}) 
and separating
the real and imaginary parts of the equations, we find
\begin{eqnarray}
\label{continuity}
\frac{\partial\,n_{i}}{\partial\,t} + 
\frac{\partial}{\partial\,x}(n_{i}u_{i}) &=& 0 \,,\\
\label{momentum}
\frac{\partial\,u_{i}}{\partial\,t} + 
u_{i}\frac{\partial\,u_{i}}{\partial x} &=& 
\frac{e}{M}\frac{\partial\phi}{\partial\,x} +
\frac{\hbar^2}{2M^2}\frac{\partial}{\partial\,x}
\left(\frac{\partial^{2}(\sqrt{n_i})/\partial\,x^2}{\sqrt{n_i}}\right) \,, \\
\label{poisson2}
\frac{\partial^{2}\phi}{\partial\,x^2} &=& 
\frac{e}{\varepsilon_0}(\sum_{i=1}^{N}n_i - n_{0}) \,.
\end{eqnarray}
Quantum effects are contained in the pressure-like, $\hbar$-dependent
term in Eq. (\ref{momentum}). If we set $\hbar = 0$, we simply obtain the
classical multistream model introduced by Dawson 
\cite{Dawson}. Therefore, 
we shall refer to Eqs. (\ref{continuity})-(\ref{poisson2}) 
as the quantum multistream model. 

Equations (\ref{continuity})--(\ref{poisson2}) constitute the
mathematical model used in the rest of this work.
We focus our attention on the one-stream (Sec. 2) and two-stream (Sec. 3)
cases, the latter being related to the well-known two-stream instability.
The relevant dispersion relations are derived, and the unstable branches
are identified. We also investigate
the properties of stationary 
solutions of the Schr\"odinger-Poisson system, which can be viewed as the 
quantum counterpart of the classical Bernstein-Greene-Kruskal 
(BGK) modes \cite{Bernstein}. 
The analytical calculations are checked against time-dependent
numerical simulations, shown in Sec. 4. 
Our conclusions are presented in Sec. 5.

\section{One-stream plasma}

In order to fix the basic ideas, we first consider the one-stream case
and take $N = 1$, i.e. a single pure quantum state. For brevity, we write $n_1
\equiv n$, $u_1 \equiv u$. We obtain
\begin{eqnarray}
\label{cont1}
\frac{\partial\,n}{\partial\,t} + \frac{\partial}{\partial\,x}(nu) 
&=& 0 \,,\\
\label{mom1}
\frac{\partial\,u}{\partial\,t} + u\frac{\partial\,u}{\partial\,x} 
&=& \frac{e}{M}\frac{\partial\phi}{\partial\,x} +
\frac{\hbar^2}{2M^2}\frac{\partial}{\partial\,x}
\left(\frac{\partial^{2}(\sqrt{n})/\partial\,x^2}{\sqrt{n}}\right) \,, \\
\label{pois1}
\frac{\partial^{2}\phi}{\partial\,x^2} &=& 
\frac{e}{\varepsilon_0}(n - n_{0}) \,.
\end{eqnarray}
A homogeneous zeroth-order solution for Eqs. (\ref{cont1})--(\ref{pois1}) is provided by
\begin{equation}
n = n_0 \,,\quad u = u_0 \,,\quad \phi = 0 \,,
\end{equation}
where $u_0$ is a constant representing the equilibrium velocity of the stream. 
The linear stability of 
this solution is obtained by Fourier analyzing Eqs. (\ref{cont1})--(\ref{pois1}),  
\begin{eqnarray}
n &=& n_0 + n'\exp(i(kx - \omega\,t)) \,, \\  u &=&  u_0 + 
u'\exp(i(kx - \omega\,t)) \,, \\ \phi &=& \phi'\exp(i(kx - \omega\,t))
\end{eqnarray}
Retaining only terms up to first order in $n'$, $u'$ and $\phi'$ leads 
to the dispersion relation
\begin{equation}
\label{disp1}
(\omega - ku_0)^2 = \omega_{p}^2 + \hbar^{2}k^{4}/4M^2 \,,
\end{equation}
where $\omega_p = (n_{0}e^{2}/M\varepsilon_{0})^{1/2}$ is the plasma frequency.
For $u_0 = 0$, the dispersion relation given in \cite{Drummond} is 
recovered, the term $ku_0$ merely representing a Doppler shift. 
As the frequency $\omega$ 
is always real, there can be neither instability nor 
damping of the wave.

The classical analog of this system is the ``cold plasma" model, which is
known to sustain nonlinear oscillations when the amplitude of the initial
perturbation is smaller than a certain value. Beyond that value, 
the solution becomes singular in a finite time, which is a sign
that the model is no longer valid. This phenomenon corresponds 
to the breaking of the plasma wave, due to particle overtaking in the phase space.
Due to the pressure-like term in Eq. (\ref{mom1}), 
the quantum solution nevers becomes singular, as 
it was shown by computer simulations \cite{Bertrand}.

Let us now turn our attention to the stationary regimes of the system. If all
quantities only depend on position, then Eqs. (\ref{cont1})-(\ref{mom1}) are
reduced to
\begin{eqnarray}
\label{cont1s}
\frac{d}{dx}(nu) &=& 0 \,,\\
\label{mom1s}
u\frac{du}{dx} 
&=& \frac{e}{M}\frac{d\phi}{dx} +
\frac{\hbar^2}{2M^2}\frac{d}{dx}
\left(\frac{d^{2}(\sqrt{n})/dx^2}{\sqrt{n}}\right) \,.
\end{eqnarray}
Equations (\ref{cont1s})--(\ref{mom1s}) possess the first integrals
\begin{eqnarray}
\label{current}
J &=& nu \,,\\
\label{energy}
E &=& \frac{Mu^2}{2} - e\phi -
\frac{\hbar^2}{2M}
\left(\frac{d^{2}(\sqrt{n})/dx^2}{\sqrt{n}}\right) \,,
\end{eqnarray}
corresponding to charge and energy conservation. The constant $E$ can be
eliminated by the global shift $\phi \rightarrow \phi + E/e$, and therefore we assume
$E = 0$ in the rest of this section. Eliminating $u$, introducing $A = \sqrt{n}$
and using Poisson's equation, we obtain
\begin{eqnarray}
\label{a1}
\hbar^{2}\frac{d^{2}A}{dx^2} 
&=& M\left(\frac{M J^2}{A^3} - 2e\,A\phi\right) \,,\\
\label{phi1}
\frac{d^{2}\phi}{dx^2} 
&=& \frac{e}{\varepsilon_0}(A^2 - n_0) \,.
\end{eqnarray}

It can be easily verified that the $J = 0$ case cannot sustain small-amplitude, periodic
solutions. Hence, we assume $J = n_{0}u_0$ with $u_0 \neq 0$ and introduce the
following rescaling
\begin{eqnarray}
\label{resc1}
x^{\displaystyle *} =  \frac{\omega_{p}x}{u_0} \,,
\quad A^{\displaystyle *} = \frac{A}{\sqrt{n_0}} \,, \nonumber\\
\label{resc2}
\phi^{\displaystyle *} = \frac{e\phi}{M u_{0}^2} \,, \quad 
H = \frac{\hbar\omega_p}{M u_{0}^2} \,.
\end{eqnarray}
We obtain, in the transformed variables (omitting the stars for simplicity of notation),
\begin{eqnarray}
\label{bgka}
H^{2}\frac{d^{2}A}{dx^2} &=& - 2\phi\,A + \frac{1}{A^3} \,, \\
\label{bgkphi}
\frac{d^{2}\phi}{dx^2} &=& A^2 - 1 \,,
\end{eqnarray}
a system that only depends on the rescaled parameter $H$, which is a
measure of the importance of quantum effects. 

Notice that the classical limit is singular, in the sense that Eq. (\ref{bgka}) degenerates
into an algebraic equation when $H = 0$. The equation for the electrostatic potential
then becomes
\begin{equation}
\label{classbgk}
\frac{d^{2}\phi}{dx^2} = \frac{1}{\sqrt{2\phi}} -1~.
\end{equation}
Equation (\ref{classbgk}) has a Hamiltonian character, 
and corresponds to the equation of motion
of a ``particle" moving in a potential $V(\phi) = \phi-\sqrt{2\phi}$. Using this
analogy, one can see that Eq. (\ref{classbgk}) has periodic solutions 
around the equilibrium $\phi = 1/2$, provided that
the initial condition satisfies $0 < \phi(x=0) < 2$. The fact that no solution exists for
sufficiently large values of the potential is easily understood. A large potential
fluctuation induces a velocity fluctuation, which can drive 
$u(x)$ far from its nominal value $u_0$. If the potential is sufficiently
strong, $u(x)$ can even vanish, but in that case the relation $nu = J =$ constant
implies an infinite density. 
This is the well-known effect of particle overtaking that occurs in the cold 
plasma model.

Going back to the quantum mechanical case, we have shown that
Eqs. (\ref{bgka})--(\ref{bgkphi}) describe the inhomogeneous QBGK 
(quantum-BGK) equilibria
of the one-component quantum plasma.  We are faced with the mathematical
problem of understanding the qualitative properties of the solutions of a
coupled, nonlinear system of two second order differential equations
depending on a parameter. Only few analytical results can be obtained, shown
in the rest of this section.

Equations (\ref{bgka})--(\ref{bgkphi}) can be put into Hamiltonian form by using
the variables
\begin{equation}
\bar{A} = iA \,, \quad \bar{\phi} = \phi/H \,.
\end{equation}
Notice that the rescaled 
amplitude $\bar{A}$ is a purely imaginary quantity. We have
\begin{equation}
\label{eq25}
d^{2}\bar{A}/dx^2 = - \partial\,U/\partial\bar{A} \,, \quad 
d^{2}\bar{\phi}/dx^2 = -
\partial\,U/\partial\bar{\phi} \,,
\end{equation}
where $U \equiv U(\bar{A},\bar{\phi})$ is the pseudo-potential
\begin{equation}
\label{eq26}
U(\bar{A},\bar{\phi}) = \frac{1}{H}(1 + \bar{A}^{2})\bar{\phi} +
\frac{1}{2H^{2}\bar{A}^2} \,.
\end{equation}
Since the equations of motion are autonomous with respect to the independent
variable $x$, the Hamiltonian formulation immediately gives the first integral
(subscripts denote differentiation)
\begin{equation}
\label{invt}
I = \frac{1}{2}( \bar{A}_x^2+ \bar{\phi}_x^2) +
U(\bar{A},\bar{\phi}) \,,
\end{equation}
which is the Hamiltonian function in transformed coordinates. Transforming back to
the original variables, one obtains the first integral for Eqs. 
(\ref{bgka})--(\ref{bgkphi}) 
\begin{equation}
\label{inv}
I = \frac{1}{2}( -{A}_x^2 + H^{-2}{\phi}_x^2) +
\frac{1}{H^2}(1 - A^2) \phi - \frac{1}{2H^2 A^2} \,.
\end{equation}
According to Liouville-Arnold theorem \cite{VIArnold}, 
an autonomous two-degree of freedom
Hamiltonian system is completely integrable if it possesses two first integrals
in involution and with compact level surfaces. Even if $I$ has not compact
level surfaces, a second constant of motion would be a strong indication of
integrability of the QBGK spatial dynamics. We have tried to find a second
constant of motion for Eqs. (\ref{bgka})--(\ref{bgkphi}) by a variety of methods
(geometrical Noether and Lie symmetries, for instance), but without success. 
However, numerical integrations of Eqs. (\ref{bgka})--(\ref{bgkphi}) for a 
wide range of values of $H$, and different initial conditions, strongly suggest 
that, when bounded solutions exist, they are always regular. An additional first
integral must therefore exist, although its actual expression may be difficult 
to guess.

It is interesting to perform a linear stability analysis in order to see in
what conditions the system supports small amplitude spatially periodic
solutions. Writing
\begin{equation}
A = 1 + A'\exp(ikx) \,,\quad \phi = 1/2 + \phi'\exp(ikx)~,
\end{equation}
and retaining in Eqs. (\ref{bgka})--(\ref{bgkphi}) only terms up to first order in
the primed variables, we obtain the relation
\begin{equation}
\label{quartic}
H^{2}k^{4} - 4 k^2 + 4 = 0 \,.
\end{equation}
Again, we point out the singular character of the classical limit: for
$H = 0$, Eq. (\ref{quartic}) degenerates into a
quadratic equation, with solutions $k = \pm 1$. The wavenumbers being always
real, this corresponds to spatially periodic solutions. When
$H \neq 0$, we obtain
\begin{equation}
k^{2} = \frac{2 \pm 2\sqrt{1 - H^{2}}}{H^2}
\,.
\end{equation}
For $H < 1$ (semiclassical regime), both wavenumbers are real, and 
therefore the system can sustain spatially
periodic oscillations. For $H > 1$ (strong quantum
effects), the solutions are spatially unstable, and grow exponentially. 
For $H = 1$, the spectrum is degenerate, with associated secular terms. 
The corresponding solution is also spatially unstable, growing linearly
with $x$.
We conclude that small-amplitude stationary solutions of the
one-stream Schr\"odinger-Poisson system can
only exist in the semiclassical regime, $H < 1$.

\section{Two-stream plasma}

\subsection{Two-stream instability}
We now turn to the more interesting case of two streams, which classically
can give rise to instability. For this, we
consider Eqs. (\ref{continuity})--(\ref{poisson2}) with $N = 2$.
We first linearize around the equilibrium solution
$n_1 = n_2 = n_{0}/2$, $u_1 
= - u_2 = u_0$, $\phi = 0$, where $u_0 \neq 0$ is a nonzero 
reference velocity, and then Fourier transform both in space and time
variables. In terms of the dimensionless variables
\begin{equation}
\label{nondim}
\Omega = \omega/\omega_p~, ~~~ K = u_{0}k/\omega_p~, ~~~
H = \hbar\omega_{p}/M u_{0}^2~,
\end{equation}
the dispersion relation becomes

\begin{equation}
\label{disp2}
\Omega^4 - \left(1 + 2K^2 + \frac{H^{2}K^{4}}{2}\right)\Omega^2 
- K^{2}\left(1 - \frac{H^{2}K^2}{4}\right)\left(1 - K^2 + 
\frac{H^{2}K^4}{4}\right) = 0 \,.
\end{equation}
Consequently, one obtains
\begin{equation}
\label{omega}
\Omega^2 = \frac{1}{2} + K^2 + {H^2 K^4 \over 4} 
\pm\frac{1}{2}(1 + 8K^2 + 4H^2 K^6)^{1/2}
\,.
\end{equation}
The solution for $\Omega^2$ has two branches, one of which
is always positive and gives stable oscillations. The
other solution is negative ($\Omega^2 < 0$) provided that
\begin{equation}
\label{cond}
(H^{2}K^{2} -4)(H^2 K^4 -4K^2 + 4 ) < 0~,
\end{equation}
Notice that instability,
when it occurs, always arises through the marginal mode ($\Omega = 0$).
In the classical case, Eq. (\ref{cond}) yields $K^2 < 1$, which is the
classical criterion for the occurrence of the 
two-stream instability. In the quantum case, Eq. (\ref{cond}) 
bifurcates for $H=1$. If $H>1$, the second factor is always positive, and the plasma is
unstable if $HK<2$. If $H<1$, there is instability if either
\begin{equation}
0 < H^2 K^2 < 2-2\sqrt{1-H^2} ~,
\end{equation}
or
\begin{equation}
2+2\sqrt{1-H^2} < H^2 K^2 < 4~.
\end{equation}
This yields the stability diagram plotted on Fig. 1. The lower instability zone
is semiclassical, as it represents an extension of the classical instability
criterion. The upper instability zone, on the other hand, has no classical analog.
The two zones merge for $H=1$. We call $K_A, K_B$ and $K_C$
the wavenumbers at which the growth rate
vanishes. These are defined by
\begin{eqnarray}
\label{kab}
H^2 K^2_{A,B} &=& 2 \pm 2\sqrt{1-H^2}~,\\
\label{kc}
H^2 K^2_{C} &=& 4~.
\end{eqnarray}
It is easy to verify the following property
\begin{equation}
K^2_{A} + K^2_{B} = K^2_{C}~.
\end{equation}
We shall see that these wavelengths are related to the stationary
solutions of the Schr\"odinger-Poisson system (quantum-BGK, or QBGK, modes).
>From the previous analysis, it appears that quantum mechanics has a 
destabilizing effect in the semiclassical regime, where more modes
are unstable compared to the classical case. On the other hand, 
when $H > 2$, fewer modes turn
out to be unstable than in the classical regime. This is the result
obtained by Suh, Feix and Bertrand \cite{Suh}, who found that quantum
effects are stabilizing for large enough $H$. 

It would be interesting to know whether the quantum instability is 
stronger or weaker than the classical one. For this, one should search
for the maximum (over all wavenumbers) growth rate for a fixed value
of $H$, and compare it to the maximum classical growth rate. Although the 
algebra is rather involved, direct inspection of the function $\gamma(K)$ for
different values of $H$ indicates that the maximum classical growth rate
is always larger than the maximum quantum growth rate (Fig. 2) 
[on this figure, the intersections with the $K$ axis correspond to 
wavenumbers $K_{A}$, $K_{B}$ and $K_{C}$ as defined in 
Eqs. (\ref{kab})--(\ref{kc}), for which the growth rate vanishes].
Of course, for some wavenumbers, 
the quantum growth rate may be larger than the classical
one, as can be seen from Fig. 2. 
Note that the secondary maximum (between wavenumbers $K_B$
and $K_C$) is considerably smaller than the one between $K=0$
and $K_A$ , and goes to zero in the classical limit $H \to 0$.

In the previous discussion, we have ignored the fact that, due to the 
periodic boundary conditions, the momentum variable $p=\hbar k$ is discrete, 
and can only be a multiple of $\hbar k_0$, where $k_0 = 2\pi/L$
is the fundamental wavenumber. For the previous stability calculations, we have
assumed an initial equilibrium solution for which the density $n_0$ is spatially uniform
and the velocity is equal to $\pm u_0$. This corresponds to wavefunctions 
of the type
\begin{equation}
\label{equil}
\psi_{\pm}(x,0) = \sqrt{n_0/2}~\exp(\pm~i M u_0 x/\hbar)~.
\end{equation}
In order to satisfy the periodicity, one must have
\begin{equation}
M u_0 = n \hbar k_0~,~~~n=1,2,3...
\end{equation}
Using the dimensionless variables introduced earlier, this condition becomes
\begin{equation}
\label{k0}
H K_0 = 1/n~,
\end{equation}
which implies $HK_0 \le 1$. This result sets an upper bound on the fundamental 
wavenumber $K_0$. On the instability diagram of Fig. 1, this means that the upper
instability region cannot be accessed for $K_0$, although of course it can for 
some of its harmonics. Computing the intersection of the curve $HK=1$ with the lower curve
on the diagram ($K=K_A$), we obtain $H^2 = 3/4$. Therefore, in the region 
$H^2 > 3/4$, the fundamental wavenumber is always unstable.

The previous discussion raises the question of the physical 
meaning of the upper instability region in Fig. 1. In particular, we ask whether it is
possible to locate the fundamental wavenumber in the stable region (between $K_A$
and $K_B$), and a higher harmonic $K_m = m K_0$ (where $m$ is an integer) 
in the unstable upper zone (between $K_B$ and $K_C$). The constraints to 
be satisfied are
\begin{eqnarray}
K_B &<& K_m < K_C~,\\
K_0 &>& K_A~.
\end{eqnarray}
Taking into account Eq. (\ref{k0}), we can write $K_m = m/(nH)$, and the previous
inequalities become
\begin{eqnarray}
\label{mn1}
2+2\sqrt{1-H^2} &<& \frac{m^2}{n^2} < 4~,\\
\label{mn2}
2-2\sqrt{1-H^2} &<& \frac{1}{n^2}  ~.
\end{eqnarray}
The square root can be eliminated by summing Eqs. (\ref{mn1})-(\ref{mn2}).
We obtain
\begin{equation}
4n^2-1 < m^2 < 4n^2~,
\end{equation}
which cannot be satisfied for any pair of integer numbers $(n,m)$.
In summary, it is not possible to excite a harmonic in the unstable upper zone
of Fig. 1, without also exciting the fundamental mode in the lower unstable
region. Therefore, at least in the semiclassical regime $H<1$, we cannot expect to
observe a ``purely quantum" instability. 
However, as remarked earlier, for $H^2 > 3/4$ the fundamental mode is 
unstable, as a result of quantum effects. 
We stress that the above restrictions are a result of the periodic boundary conditions,
and do not apply to a truly infinite plasma, for which momentum space is 
continuous. Still, periodic conditions can be relevant, for instance, to
solid state plasmas, where the periodicity is induced by the underlying ion
lattice.

\subsection{Stationary solutions -- QBGK modes}
Let us now consider the stationary states of the two-stream
Schr\"odinger-Poisson system (QBGK modes). If all quantities
are dependent only on position, Eqs. (\ref{continuity})--(\ref{momentum})
for $N = 2$ possess the first integrals
\begin{eqnarray}
\label{curr}
J_1 &=& n_{1}u_1 \,, \quad J_2 = n_{2}u_2 \,,\\
\label{en1}
E_1 &=& \frac{Mu_{1}^2}{2} - e\phi -
\frac{\hbar^2}{2M}\frac{d^{2}(\sqrt{n_1})/dx^2}{\sqrt{n_1}} \,,\\
\label{en2}
E_2 &=& \frac{Mu_{2}^2}{2} - e\phi -
\frac{\hbar^2}{2M}\frac{d^{2}(\sqrt{n_2})/dx^2}{\sqrt{n_2}} \,.
\end{eqnarray}
We are particularly interested in the case of two symmetric streams,
each carrying the same 
current (with opposite sign) and same kinetic energy. Therefore we write
\begin{equation}
J_1 = - J_2 = n_{0}u_{0}/2~, \quad E_1 = E_2 = M u_{0}^{2}/2 \,
\end{equation}
where $u_0 \neq 0$ is a nonzero reference velocity.
Let $n_1 \equiv A_{1}^2$ and $n_2 \equiv A_{2}^2$, and transform to
the dimensionless variables
\begin{eqnarray}
x^{\displaystyle *} = \omega_{p}x/u_0 \,&,& A_{1,2}^{\displaystyle *} = A_{1,2}/\sqrt{n_0}\,,\\
\label{continuity0}
\phi^{\displaystyle *} = (e\phi + E_1)/M u_{0}^2 \,&,& \quad H = \hbar\omega_{p}/M u_{0}^2 \,.
\end{eqnarray}
In these variables, the two conservation laws, Eqs. (\ref{en1})-(\ref{en2}), 
and Poisson's equation (\ref{poisson2}) take the form
of a six-dimensional dynamical system 
(we omit the stars for simplicity of notation)
\begin{eqnarray}
\label{eq1}
H^2 \frac{d^{2}A_{1}}{dx^2} &=& \frac{1}{4A_{1}^3} - 2\,\phi\,A_1 \,,\\
\label{eq2}
H^2 \frac{d^{2}A_{2}}{dx^2} &=& \frac{1}{4A_{2}^3} - 2\,\phi\,A_2 \,,\\
\label{eq3}
\frac{d^{2}\phi}{dx^2} &=& A_{1}^2 + A_{2}^2 - 1 \,.
\end{eqnarray}
Equations (\ref{eq1})--(\ref{eq3}) constitute a coupled,
nonlinear system of three second-order ordinary differential
equations, depending on the control parameter $H$.
They can
be put into a Hamiltonian form, using a procedure similar to the one 
employed for  the one-stream QBGK equations (\ref{bgka})--(\ref{bgkphi}). 
This immediately provides a first integral, which is the Hamiltonian function
itself (its actual expression is rather involved, and not particularly
illuminating). Just as in the one-stream case, we could not find any
additional constants of the motion, so that we cannot prove rigorously
that the system is integrable. However, numerical integrations of 
Eqs. (\ref{eq1})--(\ref{eq3}) (see end of this section)
strongly suggest that the system is indeed integrable, with quasi-periodic
solutions.

For the study of small amplitude oscillations, it suffices to 
expand Eqs. (\ref{eq1})--(\ref{eq3}) in the vicinity of the 
spatially homogeneous equilibrium $A_1 = 
1/\sqrt{2}$, $A_2 = 1/\sqrt{2}$, $\phi = 1/2$. 
After Fourier transforming,
the following system is obtained, for the perturbed quantities
$A'_i,~ \phi'$ 
\begin{eqnarray}
\label{eigen1}
(4-H^2 K^2)A'_i + \sqrt{2} \phi' &=& 0~,~~~i=1,2\\
\label{eigen2}
\sqrt{2} (A'_1+A'_2) + K^2 \phi' &=& 0~,
\end{eqnarray}
where $K=k u_0/\omega_p$ is the dimensionless wavenumber.
By searching for nontrivial solutions, one obtains the relation
\begin{equation}
\label{condbgk}
(H^{2}K^2 - 4)(H^{2}K^4 - 4K^2 + 4) = 0 \,,
\end{equation}
Notice that this is the same 
(with an equality sign) as the previously obtained Eq. (\ref{cond}).
Solutions of Eq. (\ref{condbgk}) represent wavenumbers for which both
the real and the imaginary part of the frequency vanish, and 
can be considered as the homogeneous limit of generally inhomogeneous
stationary states (QBGK modes).
If $H<1$ there are three such solutions, which are the wavenumbers 
$K_A, K_B$ and $K_C$ defined in Eqs. (\ref{kab})--(\ref{kc}). If $H>1$,
only the solution $K_C$ survives. The other two solutions 
become complex, so that spatially 
periodic QBGK modes can no longer exist.

It is also interesting to look for the eigenvectors corresponding to
the eigenvalues $K_A, K_B$ and $K_C$. From Eqs. (\ref{eigen1})--(\ref{eigen2})
we can express the perturbed amplitudes $A'_i$ in terms of the potential.
Two cases are possible : (1) if $K = K_C = 2/H$, then one must have
$\phi'=0$ and $A'_1=-A'_2$ ; (2) otherwise, if $K = K_A$ or $K_B$, 
we have $A'_1 = A'_2 = \sqrt{2}\phi'/(H^{2}K^2 - 4)$.
The first case is particularly interesting. It means that the mode
characterized by wavenumber $K_C$ is a ``quasi-neutral" mode, in the
sense that the associated electrostatic potential is zero to first order. 
Indeed, one can
see that no plasma parameters (such as the plasma frequency) enter the 
definition of $K_C \equiv 2/H$. These modes can be actually accessed, for example
by choosing the fundamental wavenumber $K_0=1/H$, which is the largest 
admissible value for $K_0$ [see Eq. (\ref{k0})]. In this case, the harmonic 
$2K_0 = 2/H$ corresponds to the quasi-neutral mode.

Numerical integration of Eqs. (\ref{eq1})--(\ref{eq3}) confirms the 
previous results. For instance, it
was verified that periodic solutions only exist for $H<1$.
We take $H=0.7$ and initialize the amplitudes and
the potential (at $x=0$) with their equilibrium value, plus a small perturbation
$\epsilon$, i.e.
$\phi(0) = 1/2 + \epsilon_{\phi},~A_i(0)=(1+\epsilon_i)/\sqrt{2}$. 
In agreement with the 
discussion of the previous paragraph, if we choose $\epsilon_\phi=0$ and 
$\epsilon_1 = -\epsilon_2$, the wavenumber $K_C\simeq 2.857$ is 
linearly excited and thus dominates (Fig. 3), while 
the potential remains very small. 
On the other hand, if $\epsilon_1 = \epsilon_2$ and $\epsilon_\phi$ 
arbitrary, modes
$K_A \simeq 1.08$ and $K_B \simeq 2.645$ are linearly excited (Fig. 4). 
For generic perturbations, all three
wavenumbers are excited.
Of course these results are strictly valid only for infinitesimally 
small perturbations.
For moderate values, other modes appear (visible on Figs. 3-4, for which
$\epsilon = 0.02$), although the linear wavenumbers are still dominant.
For even larger perturbations, bounded solutions no longer exist.

\section{Time-dependent numerical simulations}

A standard numerical technique has been employed in order
to integrate the time-dependent Schr\"odinger-Poisson system.
Let us write the Schr\"odinger equation as
\begin{equation}
\label{schronum}
i\hbar\frac{\partial\psi}{\partial t} = K \psi + \Phi \psi \,,
\end{equation}
where $K$ is the kinetic part of the Hamiltonian, and $\Phi(x,t)$ is 
the potential. The Hamiltonian is split into these two parts, and each is treated
separately. For the potential part, the solution is trivial
\begin{equation}
\label{pot}
\psi^{n+1} = \psi^n \exp(-i \Phi \Delta t/\hbar)~,
\end{equation}
where $\psi^n \equiv \psi(n\Delta t)$, and $\Delta t$ is the timestep.
For the kinetic part, we use a Crank-Nicolson scheme, which is exact
to second order in $\Delta t$,
\begin{equation}
\label{kin}
i\hbar\frac{\psi^{n+1}-\psi^n}{\Delta t} = 
\frac{1}{2}(K \psi)^{n+1} + \frac{1}{2}(K \psi)^{n}~.
\end{equation}
The kinetic operator $K$ is spatially discretized by using the standard 
centered differences formula. The time evolution is obtained by subsequently
applying the potential and kinetic steps described above. Poisson's equation
is solved with a Fast Fourier Transform
technique just before the potential step (notice that the kinetic step does not
alter the spatial density, and therefore the potential).
The resulting numerical scheme is unconditionally stable and 
second order accurate in both space and time variables.
Another crucial property of the scheme is that it conserves {\it exactly} the
integral $\int |\psi|^2 dx$ \cite{Manfredi2}.
A typical resolution used in the simulation is $N_x = 512$ points, and timestep
$\Delta t = 0.02$.

The initial condition for the simulations is obtained by applying
a sudden sinusoidal potential to the 
equilibrium wavefunctions given in Eq. (\ref{equil}). In dimensionless
variables, we have 
\begin{equation}
\psi_{\pm}(x,0) = 2^{-1/2} \exp(\pm~i n K_0 x)~
\exp(i \epsilon H^{-1} \cos(K_m x))~,
\end{equation}
where $\epsilon$ and $K_m=mK_0$ are the amplitude and the
wavenumber of the perturbation, and $n,m$ are integer numbers.
We remind that one must have $H K_0 = 1/n$. 

Several simulations have been run in order to compare with the analytical results
obtained in the linear regime, with excellent agreement between the two.
As an example, we use the parameters $H=0.25$ and $n=5$ ($K_0=0.8$), and perturb
the fundamental mode ($m=1$). Figure 5 shows the evolution of two
modes of the electrostatic potential. The straight line corresponds to the linear
growth rate for $K_0$, 
as computed from Eq. (\ref{omega}), and closely matches the measured
growth rate. At saturation, several modes are present. 

The total momentum distribution $F(p)$ is given by 
the sum of the square modulus of the Fourier transform
of each wavefunction, with $p=\hbar k$. 
As pointed out earlier, momentum space is discrete,
with $\Delta p = \hbar k_0$ (= 0.2 in the above case). 
For our simulations, the total momentum 
distribution of the unperturbed wavefunctions is simply
\begin{equation}
F(p) = {1 \over 2}\delta(p-M u_0) + {1 \over 2} \delta(p+M u_0)~,
\end{equation}
where $\delta$ is the Dirac delta function.
During the linear phase, the momentum distribution remains virtually 
unchanged, whereas at saturation we observe a significant spreading in momentum space,
which extends to 
$p \simeq \pm 2.5 Mu_0$ (Fig. 6). This is similar to the behavior of the
classical two-stream instability.

As was shown in the preceding section, one cannot excite a mode in the
unstable upper region of Fig. 1, without also exciting an unstable mode in the
lower zone. However, if only the larger wavenumber is initially perturbed,
one can hope to see it grow with the correct rate before other unstable modes can
be excited. In order to do so, we take $H=0.9$ and $n=5$. The fundamental wavenumber is 
thus $K_0=2/9$, and is of course unstable since $K_A=1.18$ in this case. We only
perturb the harmonic $9K_0=2$, which falls within the upper unstable zone, since
$K_B=1.883$ and $K_A=2.222$. Of course, several other modes are also unstable (namely, those
from $K_0$ to $5K_0$), but they are not initially perturbed. 
Figure 7 shows the evolution of mode $9K_0$, which closely agrees with the 
result of the linear calculation for the growth rate. Several other linearly unstable
modes appear at a later time, so that at saturation the spectrum is large. This is
reflected in the momentum distribution, plotted on Fig. 8.
 
The results for the fully quantum regime, and particularly for $H \simeq 1$, 
are more surprising. As an example, let us take $H=1$, $K_0=1$, and perturb the
fundamental mode itself. Instability occurs as expected with the correct growth
rate (see Fig. 9). However, instead of saturating at a certain level, the system 
appears to decay with the same rate, and then to grow again. These nonlinear periodic
oscillations do not damp even for very long times, as it has been checked numerically.
Higher harmonics, which are linearly stable, are driven by the fundamental mode, and show
a similar pattern, although at a lower level. The period of the oscillations is not
universal, and depends on the amplitude of the initial perturbation, which confirms
that this is indeed a nonlinear effect. 
A similar, although less pronounced behavior, is also observed for $H>1$.
The momentum distribution is virtually unchanged over 
the entire duration of the simulation. 
It must be noted that, for parameters such that $HK_0=1$,
the minimum nonzero value of the momentum is 
$\vline p_{\rm min} \vline =M u_0$, and is therefore equal to the 
momentum of the unperturbed streams. 
In other words, the streams occupy the lowest possible
level in momentum space. 
Although we do not have a detailed explanation for this phenomenon, it appears
to be an example of a completely reversible 
quantum system, in which the initial condition is
almost perfectly reconstructed after one period \cite{Manfredi}. 
This is to be compared with the inherently
irreversible classical dynamics, for which returning to the 
initial state after saturation 
is virtually impossible. 

Finally, we have studied the evolution of perturbed QBGK equilibria.
It is particularly instructive to consider the case where only the $K_C=2/H$ mode
is present. This, as detailed in the previous section, is a ``quasi-neutral" mode,
in the sense that the associated electrostatic potential is zero to first order in
the perturbation parameter. We construct a weakly inhomogeneous equilibrium by using the
form of Eq. (\ref{as}) for the wavefunctions $\psi_{1,2}$. Both the amplitude 
and the phase should be sinusoidal functions. For the amplitudes, we have 
\begin{equation}
\label{ampli}
A_{1,2} =  \frac{1}{\sqrt{2}} \left(1 \pm \epsilon \cos (K_C x) \right)~,
\end{equation}
where $K_C = mK_0$ is the wavenumber of the QBGK mode; $m$ therefore represents the
number of density oscillations. The phases $S_{1,2}$ are obtained from the definition of 
Eq. (\ref{nu}), and remembering that,
in dimensionless units, $u_i n_i = \pm 1/2$. One obtains, to first order in 
$\epsilon$
\begin{equation}
\label{phase}
S_{1,2} = \pm~ x - 2 ~\epsilon ~K_{C}^{-1} \sin(K_C x) ~.
\end{equation}
By virtue of the relations $n H K_0=1$ and $K_C=mK_0=2/H$, we obtain that $m=2n$ must be
an even number. 

As an example, we take $H=0.5, K_C=4$. With $n=1$, we obtain $K_0=2$, and the number of spatial
oscillations is $m=2n=2$. Notice that, in this case, the fundamental wavenumber is stable.
If we had chosen $n=2$, we would have had $K_0=1$, which is unstable. With this value of
$H$, it is therefore possible to construct a stable QBGK mode displaying at most two 
spatial oscillations. In order to have more oscillations, a smaller value of $H$ should be used.
For small values of the perturbation parameter $\epsilon$, the simulations confirm
the linear analytical results, and virtually no evolution is observed; besides, the potential 
fluctuations stay small compared to the density ones.
For larger values of the perturbation ($\epsilon=0.1$), some temporal variations are
observed (Fig. 10), since the wavefunctions given 
by Eqs. (\ref{ampli})--(\ref{phase}) no longer represent
an exact stationary state. However, the periodic structure is not destroyed, and the potential
fluctuations remain an order of magnitude smaller than the density fluctuations.
These results show that a quantum plasma can support almost stationary, quasi-neutral,
periodic solutions. These display significant density fluctuations for each stream, 
but small potential fluctuations. They have no analog in a classical
two-stream plasma.

\section{Conclusion}

In this work, we have introduced a quantum multistream model to describe some 
physical phenomena arising in quantum plasmas. 
The quantum multistream model may be considered as a discrete version 
of the Wigner-Poisson model, in the same sense as the classical 
multistream model is a discrete form of the Vlasov-Poisson system. 
Indeed, it is well known \cite{Markow} that the Wigner-Poisson system 
is formally equivalent to an infinite set of Schr\"odinger 
equations, coupled by a scalar potential obeying Poisson's equations. 
However, it is often more appropriate to work with the hydrodynamic 
formulation of quantum mechanics, since it makes direct use of the same
physical quantities that are employed in classical physics
(density, velocity, pressure). Moreover, the stability analysis and 
perturbation calculations 
become straightforward in the hydrodynamic formulation. 
On the other hand, we have used the Schr\"odinger representation for the 
time-dependent simulations, since accurate numerical techniques 
for this equation are well-known from the computational literature.

For the case of the two-stream instability, it has been
shown that the dispersion relation possesses three branches, one of a 
semiclassical character and two of a purely quantum nature. 
It is interesting to observe that even the classical 
branch reveals some unexpected
features: for small $H$, quantum effects tend to enhance the 
two-stream instability. More precisely, some classically stable wavenumbers 
are destabilized for sufficiently large values of $H<1$. On the other hand,
strong quantum effect can yield the opposite result:
for $H > 2$, some classically unstable wavenumbers become stable. 
The purely quantum region of the dispersion relation (the upper
unstable zone of Fig. 1) cannot however be excited without also exciting some
wavenumber in the lower (semiclassical) region. This means that a purely
quantum instability cannot be observed for $H<1$. Extensive
numerical simulations, run for different values of the relevant parameters,
wholly support the analytical results obtained from linear theory.
In the fully quantum case, we have observed a surprising,
yet unexplained, regime of undamped
nonlinear oscillations. This is a purely quantum effect, which is probably linked
to quantum recurrences and echoes \cite{Manfredi}.

We have also considered the stationary states of the 
Schr\"odinger-Poisson system, which can be viewed as the quantum analog
of the classical BGK modes. Such QBGK modes are described by a 
nonlinear system of coupled second-order differential equations,
parametrized by the dimensionless Planck's constant $H$. Such system provides an 
adequate framework for the analysis of QBGK modes, a 
task which would be rather difficult in the Wigner-Poisson formalism.
In particular, numerical simulations have shown that quasi-neutral, spatially
periodic, stationary states can be created in the two-stream plasma, 
and can survive over long times.

Some interesting questions remain to be addressed. 
In the present work, we have considered in detail only the one and 
two-stream cases. Further investigations are needed 
to explain the properties of the quantum multistream model when the
number of streams is large. In this case, the Wigner-Poisson system 
could be a more appropriate model, despite its intrinsic mathematical difficulties.
However, an intermediate number of streams might be sufficient to describe the 
main physical phenomena. Linear and nonlinear Landau damping, for example, should be 
good candidates to test these ideas, both analytically and numerically.

{\bf Acknowledgments}

We would like to thank Pierre Bertrand for his valuable comments 
and suggestions. 
One of us (F. H.) thanks the Laboratoire de 
Physique des Milieux Ionis\'es for the hospitality while this work 
was carried out, and the Brazilian agency Conselho Nacional de Desenvolvimento
Cientifico e Tecn\'ologico (CNPq) for financial support.

\newpage
\begin{center}
{\bf Figure Captions}
\end{center}

{\bf Figure 1} : Stability diagram for the two-stream plasma. The filled area 
is unstable. The dashed line corresponds to $HK=1$. 
Lower and middle solid curves: $K_{A}^2$ and $K_{B}^2$ as defined in Eq. (\ref{kab}). 
Upper solid curve: $K_{C}^2$ as defined in Eq. (\ref{kc}).

{\bf Figure 2} : Plot of the squared growth rate $\gamma$ as a function of the
dimensionless wavenumber $K$, for different values of Planck's constant.
$H=0.5$, solid line; $H=1$, dashed line;  $H=2$, dotted line. 
The intersections of these curves with the $K$ axis correspond to 
wavenumbers $K_{A}$, $K_{B}$ and $K_{C}$ as defined in 
Eqs. (\ref{kab})--(\ref{kc}).
For $H=0.5$, only the intersection at $K_A \simeq 1.035$ is shown;
the intersections at $K_B \simeq 3.864$ and $K_C=4$ are outside the $K$ axis range. 
For $H=1$, the intersections are at $K_{A}=K_{B}=\sqrt{2}$ and $K_C=2$. 
For $H=2$, there is only one intersection at $K_C=1$. 

{\bf Figure 3} : Stationary solution of the two-stream Schr\"odinger-Poisson
plasma (QBGK mode), with $H=0.7$, $\epsilon_\phi=0,~\epsilon_1= -\epsilon_2=0.02$. 
(a) Spatial variation of the density fluctuations $A'_1$ (solid line),
$A'_2$ (dashed line), and potential fluctuations $\phi'$ (dotted line).
Notice that the potential remains small.
(b) Fourier transform of $A'_1$: the linear wavenumber $K_C \simeq 2.857$ is 
dominant.

{\bf Figure 4} : Stationary solution of the two-stream Schr\"odinger-Poisson
plasma (QBGK mode), with $H=0.7$, $\epsilon_\phi=0,~\epsilon_1= \epsilon_2=0.02$. 
(a) Spatial variation of the density fluctuations $A'_1$ (solid line),
$A'_2$ (dashed line), and potential fluctuations $\phi'$ (dotted line). 
Note that the solid and
dashed lines are superposed, since $A'_1 \simeq A'_2$.
(b) Fourier transform of $A'_1$: the linear wavenumbers $K_A \simeq 1.080$ 
and $K_B \simeq 2.645$  are dominant.

{\bf Figure 5} : Two-stream instability -- evolution of the fundamental mode $K_0=0.8$
(solid line), and first harmonic $2K_0$ (dashed line) of the electrostatic
potential, for $H=0.25$. The straight line corresponds to the growth rate of $K_0$ computed
from linear theory, $\gamma = 0.3116$.

{\bf Figure 6} : Momentum distribution for the same case of Fig. 5, at times
$t=0$ (dotted line) and $t=80$ (solid line). The final distribution has been
magnified by a factor two. Momentum space is discrete with $\Delta p =\hbar k_0 =0.2$.

{\bf Figure 7} : Two-stream instability -- evolution of mode $9K_0$ of the electrostatic
potential, for $H=0.9$ and $K_0=2/9$. 
The straight line corresponds to the growth rate computed 
from linear theory, $\gamma = 0.1085$.

{\bf Figure 8} : Momentum distribution for the same case of Fig. 7, at times
$t=0$ (dotted line) and $t=160$ (solid line). The final distribution has been
magnified by a factor two. Momentum space is discrete with $\Delta p =\hbar k_0 =0.2$.

{\bf Figure 9} : Two-stream instability -- evolution of the fundamental mode 
$K_0=1$ (solid line), and first harmonic $2K_0$ (dashed line) of 
the electrostatic potential, for $H=1$.
The straight line corresponds to the growth rate of mode $K_0$ computed 
from linear theory, $\gamma = 0.2297$.

{\bf Figure 10} : Evolution of a strongly nonlinear ($\epsilon=0.1$) 
quasi-neutral ($K=K_C=4$) QBGK equilibrium, for the set of
parameters: $H=0.5, K_0=2$. The number of spatial
oscillations is $m=2$.  The density fluctuations $n'_{1}$ (solid line) 
and $n'_{2}$ (dashed line), and potential fluctuations $\phi'$ (magnified by a
factor 100, dotted line), are shown at different times.

\end{document}